\documentclass{ws-procs975x65}

\begin{document}

\title{INVESTIGATION OF LOW-DENSITY SYMMETRY ENERGY VIA NUCLEON AND FRAGMENT OBSERVABLES}

\author{Hermann H. Wolter}

\address{Fak. f. Physik, University of Munich,\\
D-85748 Garching, Germany\\
$^*$E-mail: hermann.wolter@lmu.de}

\author{J. Rizzo, M. Colonna, M. Di Toro, V. Greco}

\address{Lab. Nazionali del Sud, INFN,\\
I-95123 Catania, Italy}

\author{V. Baran}

\address{Univ. of Bucharest and NIPNE-HH,\\
Bucharest, Romania}

\author{M. Zielinska-Pfabe}

\address{Physics Dep., Smith College,\\
Northampton, Mass., USA}

\begin{abstract}
With stochastic transport simulations we study in detail central and peripheral collisions at Fermi energies and suggest new observables, sensitive to the symmetry energy below normal density.
\end{abstract}


\bodymatter

\vspace{1cm}

There has been much interest in recent years in
the determination of the nuclear symmetry energy as a function of density,
which is important for the structure of exotic nuclei as well as for astrophysical processes.
Heavy ion collisions (HIC) present an attractive way to constrain the
the existing models for this poorly-determined isovector equation-of-state (iso-EOS) \cite{fuchswci},
which can be investigated both at densities above and below
normal density with relativistic energies and in the Fermi energy domain, respectively. However, observables, which
are both sensitive to the iso-EOS and testable
experimentally, still have to be identified  clearly \cite{baranPR,Isospin01}.

In this report we discuss dissipative collisions at
Fermi energies.
Isospin dynamics at low and intermediate energies and its
relation to the symmetry
energy  has, in fact,  attracted much attention in recent years in  experiment
as well as in theory \cite{isotr05,tsang92,shiPRC68,BALi}.
Here we  focus our attention on pre-equilibrium emission in central collisions and on the charge equilibration dynamics in peripheral collisions, where
we expect to see symmetry energy effects. The interesting feature  at  Fermi energies is the onset of collective flows
due to compression and expansion of the interacting nuclear matter. The
isospin transport takes place in regions with   density and asymmetry variations, and thus
we expect to have contributions  to the isospin current from charge and mass
drift mechanisms.

\def\figsubcap#1{\par\noindent\centering\footnotesize(#1)}
\begin{figure}[t]%
\begin{center}
 \parbox{5.5cm}{\epsfig{figure=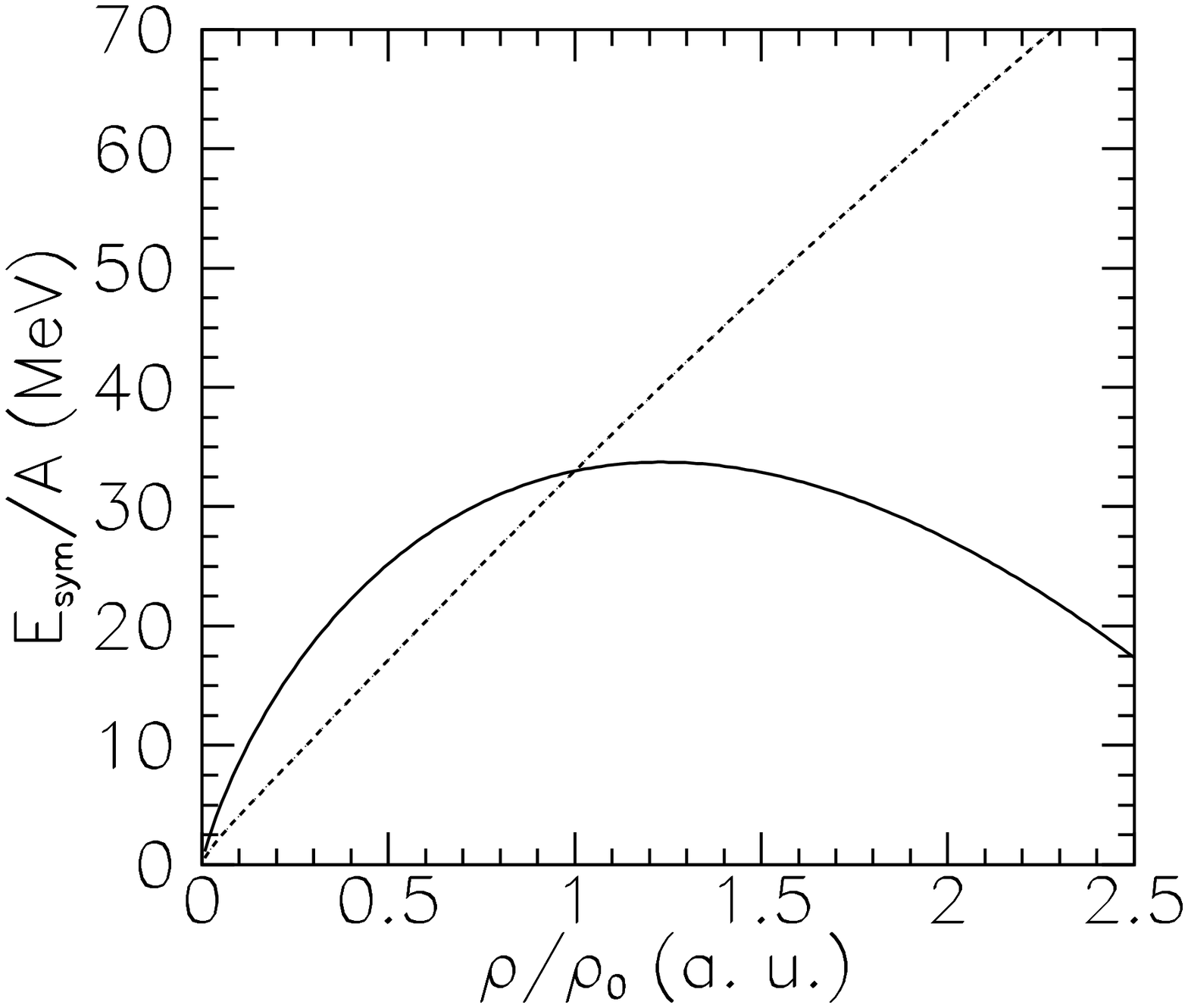,width=5.3cm}
 \figsubcap{a}}
 \hspace*{0.5cm}
 \parbox{6cm}{\epsfig{figure=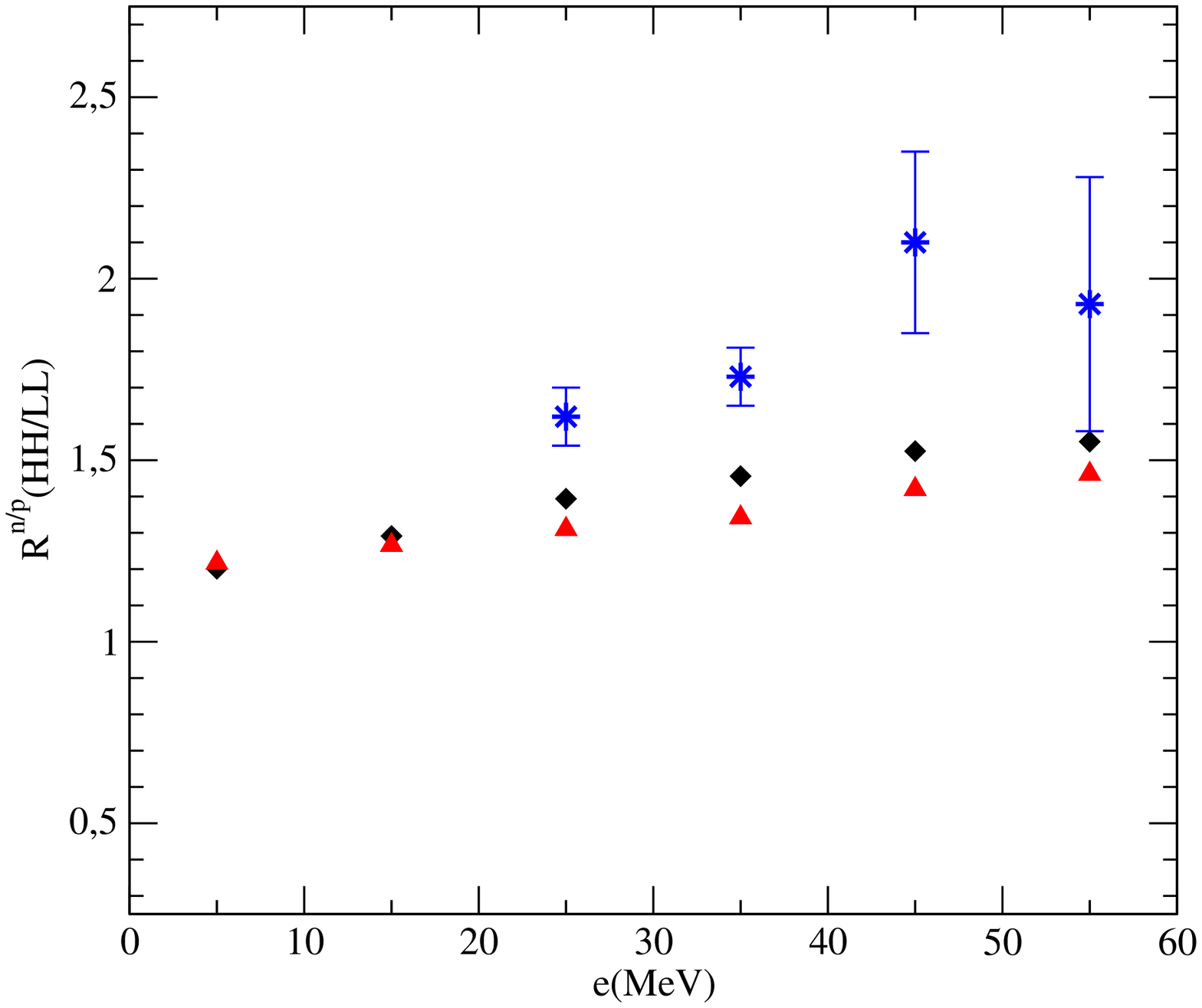,width=5.5cm,
 angle=-90}
 \figsubcap{b}}
 \caption{
 (a) Density dependence of the  symmetry energies  used in the simulations
presented here:  asy-soft (solid) and asy-stiff (dashed).
\newline
 (b)Double ratio of emitted neutron over proton yield in central $^{124}Sn+^{124}Sn$
 (HH) over $^{112}Sn+^{112}Sn$ (LL) collisions at 50 $AMeV$ with asy-stiff
 (red triangles) and asy-soft (black diamonds) EOS as a function of nucleon kinetic energy. Data of Famiano et al., \cite{famiano} are given as (blue) stars.}
\label{eos_DR}
\end{center}
\end{figure}

We perform {\it ab initio} collision simulations using
the microscopic Stochastic Mean Field (SMF) model.
It is based on mean field
 transport theory  with correlations from hard nucleon-nucleon
collisions and  with stochastic fluctuations acting on the mean
phase-space trajectory
\cite{fabbri04,chomazPR}.  Stochasticity is essential in
order   to allow the growth of dynamical
instabilities with fragment production, and to obtain physical widths
of distributions of observables.  A detailed description of the procedure is given in ref. \cite{baranPR} and in refs. therein.

We have used
a  generalized form of effective interaction  with momentum dependent terms  in the isoscalar and the
isovector channel  \cite{rizzoPRC72}, which is
 an asymmetric extension of the Gale-Bertsch-DasGupta  (GBD)  force
 \cite{GalePRC41}.
The parameters are chosen to give a soft equation of state
for $symmetric$ nuclear matter
(compressibility modulus 215 MeV, isoscalar effective mass $m^*/m=0.67$),  which is held fixed.
Here we want to test the sensitivity of isospin transport observables
to two essentially different behaviors of the symmetry energy around saturation: {\it asy-soft} and {\it asy-stiff} \cite{baranPR}.
In Fig.1a we show the density dependence for these two typical choices.
The {\it isoscalar} momentum dependence has been found to be important for the general dynamics, in particular the particle flow, of HIC \cite{DanLac02}. To discuss its influence also on isospin dynamics we consider here interactions with (MD) and without (MI) momentum dependence. The {\it isovector} momentum dependence changes the proton/neutron effective masses and is still very controversial \cite{fuchswci,baranPR}. It is most effective at higher energies, but it effects are also evident in the Fermi energy range in pre-equilibrium emission\cite{rizzoPRC72}.

Isospin transport is closely connected to the value and the slope of the symmetry energy at a given density.
In fact,  the $p/n$ currents can be expressed as
$${\bf j}_{p/n} = D^{\rho}_{p/n}{\bf \nabla} \rho - D^{\beta}_{p/n}{\bf \nabla}
 \beta $$
with $D^{\rho}_{p/n}$ the  mass (drift), and
 $D^{\beta}_{p/n}$  the  isospin transport (diffusion) coefficients (asymmetry $\beta=(N-Z)/A$), which are directly
given by the $n,p$ chemical potentials \cite{isotr05}.
Of special interest here is the difference  of
neutrons and protons currents (iso-vector current)
for which the transport coefficients are proportional to \cite{isotr05}
\begin{eqnarray}
D^{\rho}_{n} - D^{\rho}_{p}  & \propto & 4 \beta \frac{\partial E_{sym}}
{\partial \rho} \, ,  \nonumber\\
 D^{\beta}_{n} - D^{\beta}_{p}  & \propto & 4 \rho E_{sym} \, .
 \nonumber
\end{eqnarray}
Referring back to Fig. 1a we see that isospin drift and diffusion behave very differently for the two iso-EOS's.

Preequilibrium nucleons and light clusters are emitted in the approach and overlap stages of a HIC. The ratio of neutron to proton yields, (resp. of isobaric light cluster yields) carries information on the isospin forces. To reduce effects of secondary emission double ratios between different reactions have been investigated for nucleons \cite{famiano,danielew} and IMF's \cite{mc_DR}. We show a result from our calulations for nucleons in the ``gas'' phase (defined by a density cut of $\rho/\rho_0<1/6$) in Fig. 1b. It is seen that the asy-soft EOS is more effective, since the symmetry energy is higher below normal density (see Fig. 1a).

However, the iso-EOS effect is not very large, and both results are considerably below the data \cite{famiano}, which, unlike in our calculations, are taken with a transverse angular cut. Already the results for the single $n/p$ yield ratios deviate strongly from the experiment, which are very much higher for low energy nucleons. Our result is also in contrast to other calculations \cite{famiano,danielew}, which show a stronger iso-EOS effect. Obviously, both the experimental data and the calculations have to be understood better. We also mention, that the effect of different neutron/proton effective masses, i.e. of an isovector momentum dependence, is of considerable influence already in this energy range \cite{rizzoPRC72}, but does not resolve the discrepancies noted above. The calculations shown in Fig. 1b are taken for the choice $m^*_n>m^*_p$.

\def\figsubcap#1{\par\noindent\centering\footnotesize(#1)}
\begin{figure}[t]
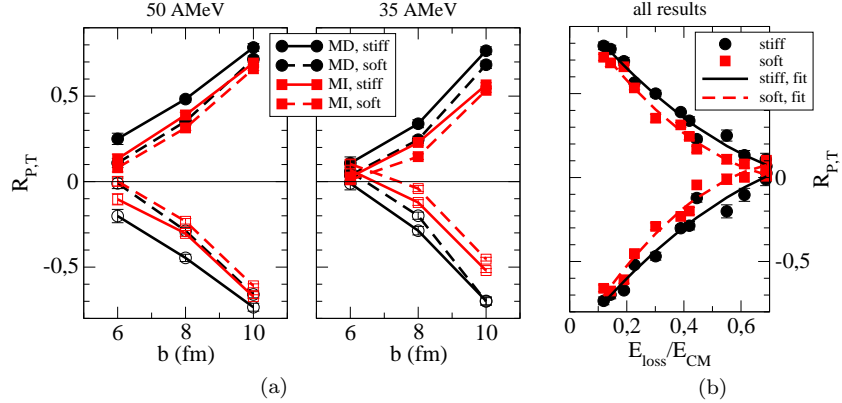
%
\begin{center}
 \parbox{7cm}{\epsfig{figure=wolter3.eps,width=6.8cm}
 \figsubcap{a}}
 \hspace*{0.1cm}
 \parbox{4cm}{\epsfig{figure=wolter4.eps,width=3.7cm}
 \figsubcap{b}}
 \caption{
 (a) Imbalance ratios for $Sn + Sn$ collisions for incident energies
of 50 (left)
and 35 $AMeV$ (right) as a function of the impact parameter
for asy-stiff and asy-soft EOS (signatures see legend box),
 projectile rapidity (upper curves), target rapidity (lower curves).
\newline
 (b)Imbalance ratio for all results in part (a) but
 as a function of relative energy loss for asy-stiff
 (black dots) and asy-soft (red squares) EOS.
Quadratic fit to all points for the asy-stiff (solid), resp.
 asy-soft (dashed)  EOS.}
\label{imb_ratio}
\end{center}
\end{figure}

In peripheral collisions of nuclei with different asymmetries isospin is equilibrated through the neck, mainly due to isospin gradients (diffusion). The amount of isospin transport has been measured with the so-called imbalance (or isospin transport) ratio
\cite{ramimb}, which is defined as
$$R^{\beta}_{P,T} = 2\frac{\beta^M-\beta^{eq}}{\beta^{HH}-\beta^{LL}} \, ,$$
with $\beta^{eq}=(\beta^{HH}+\beta^{LL})/2$.
Instead of the asymmetry $\beta=(N-Z)/A$,
one has also considered other isospin sensitive quantities, such as isoscaling coefficients \cite{WCI_betty}.
The indices $HH$ and $LL$ refer to the symmetric reaction
between the
heavy  ($n$-rich,$^{124}Sn$) and the light ($n$-poor,$^{112}Sn$)  systems, while $M$ refers to the
mixed reaction, and
$P,T$ denote the
PLF and TLF rapidity regions. Clearly, this ratio is $\pm1$ for complete transparency, resp. complete
rebound, while it is zero for complete equilibration.  Indeed, it can be shown, that the imbalance ratio depends on the magnitude of the symmetry energy and the interaction time. It is a very sensitive observable, magnifying small differences in asymmetry.

Results for our system as a function of impact parameter for different beam energies are shown in Fig. 2a. It is seen that the equilibration is larger (R smaller) for an asy-soft EOS (as expected from above), and for MI interactions and lower energy. The last two observation are fairly obvious, since they are due to longer interaction times (since the collision is also faster for the repulsive MD forces). It is therefore profitable to consider the imbalance ratio as a function of the interaction time, or an observable which is closely correlated to it. Such an observable is the total kinetic energy loss, which has been extensively investigated in dissipative collisions \cite{soul04}. We thus show the imbalance ratios as a function of the relative energy loss per particle in Fig. 2b. Here all results for MD/MI interactions and 35/50 $AMeV$ are collected and - to guide the eye - are fitted by a quadratic curve for the asy-stiff and asy-soft EOS's separately. It is now seen that the results presented in this way are {\it only} sensitive to the iso-EOS, albeit with some scatter.  Such a representation should be useful to obtain a unifying picture of different experiments as well as  calculations.

\def\figsubcap#1{\par\noindent\centering\footnotesize(#1)}
\begin{figure}[t]
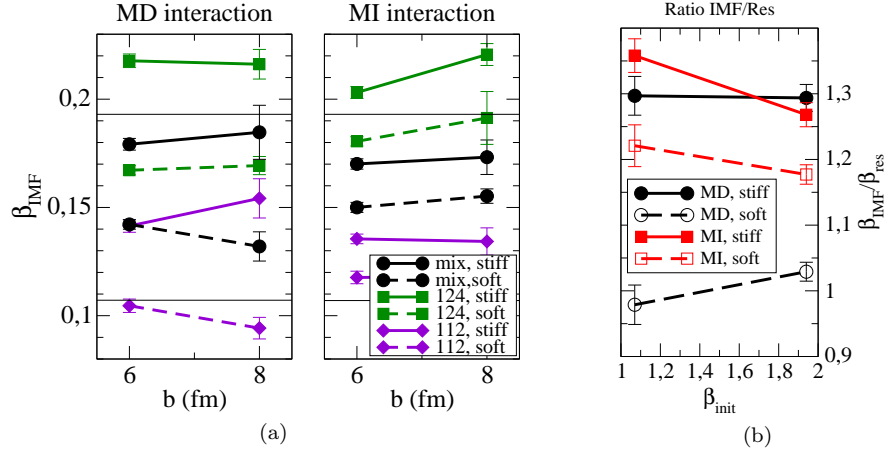
%
\begin{center}
 \parbox{7cm}{\epsfig{figure=wolter5.eps,width=6.8cm}
 \figsubcap{a}}
 \hspace*{0.8cm}
 \parbox{3.8cm}{\epsfig{figure=wolter6.eps,width=3.6cm}
 \figsubcap{b}}
 \caption{
 (a) Asymmetries of IMF's in ternary $Sn+Sn$ reactions at 50 AMeV
as a function of impact parameter for MD (left panel) and MI (right panel)
interactions for mixed and symmetric $Sn+Sn$ collisions for asy-stiff and
asy-soft EOS's (see legend box). Horizontal thin
lines: asymmeties
of $^{124}Sn$ and $^{112}Sn$, respectively. \newline
 (b) Ratios of asymmetries
of IMF to residues for symmetric $Sn+Sn$
reactions at $50 AMeV$ as a function of the initial asymmetry  for
$b=6fm$ for MD and MI interactions (see legend box).
}
\label{bet_ratio_IMF}
\end{center}
\end{figure}

In peripheral collisions a third intermediate mass particle (IMF) can appear, which originates in the rupture of the neck (ternary event, neck fragmentation)\cite{MDT_WCI}. This has been studied also experimentally in asymmetric collisions, in particular with respect to velocity correlations between the residues and the IMF and to the alignment of the IMF \cite{DeFilipp05}. In this report the isospin content of the IMF is of special interest, since it is mainly influenced by isospin transport due to density gradients, i.e. isopin drift or migration, which according to the above is governed by the slope of the symmetry energy below normal density. The asymmetry of the IMF in ternary reactions in our systems is shown in Fig. 3a both for the symmetric and the mixed $Sn+Sn$ collisions.

The asymmetry of the IMF is larger, i.e.
the IMF is more n-rich, for the stiff  relative to the soft iso-EOS,
since the former  exhibits a larger isospin migration due to
the larger slope of the symmetry energy below saturation. This is clearly the case for the
symmetric reactions, but it is also true for the mixed reactions, where there is a competition with
isospin diffusion, which depends on the value of the symmetry
energy and it is larger for the
soft iso-EOS. Our result then shows that the isospin migration is the
dominating effect for the asymmetry of the neck fragments.
It is also seen that the difference between stiff and soft iso-EOS is particularly large for MD interactions, which can be traced back to the fact that the IMF's originate from more compact configurations.

The sensitivity to the IMF asymmetry can be enhanced by taking the ratio relative to the  asymmetry of the residue, which is shown (only for the more transparent symmetric reactions) in Fig. 3b, as a function of the initial asymmetry. The large, almost $30\%$ effect for the more realistic MD interaction is noteworthy. Thus this quantitiy, which should also not be very sensitive to secondary evaporation, may constitute a promising observable to gain more information on the symmetry energy. One may even consider double ratios of this quantity in reactions of HH over LL $Sn$ isotopes.

We have shown that there exist several observables which should be able to yield information on the ill-determined low-density symmetry energy. Here we have investigated ratios of pre-equilibrium particles in central collisions and isospin transport between the residues and to the neck in peripheral collisions. We suggest to study the imbalance ratio not only as a function of centrality, but also depending on the energy loss in the reaction. We have also identified the asymmetry of an IMF from the neck as a promising observable. Unfortunately, however, both the agreement of theoretical calculations with each other  as well as the comparison with experimental data are still far from satisfactory, such that the question of the isovector EOS has to be considered still open. Generally, the investigations into the low density iso-EOS favor an asy-stiff EOS from the isospin diffusion and the neck fragmentation data, and rather an asy-soft behavior from the pre-equilibrium studies. Further intensive work, perhaps also with data from more asymmetric radioactive beams, is highly desirable.

This work has been supported by the BMBF, Germany, grant 06LM189, by the DFG Cluster of Excellence {\it Origin and Structure of the Universe}, and by the Romanian Min. of Educ. and Research, contract CEX-05-D10-02.


\begin{thebibliography}{99}

\bibitem{fuchswci}
 C.Fuchs, H.H.Wolter,
{\em Eur. Phys. Jour.} {\bf A30} 5 (2006), and refs. therein.

\bibitem{baranPR} V. Baran, M. Colonna, V. Greco, M. Di Toro, {\em Phys. Rep.}
{\bf 410} (2005) 335.

\bibitem{Isospin01} {\it Isospin Physics in Heavy-ion Collisions at
Intermediate Energies}, Eds. B.A. Li and W. Udo Schr\"oder, Nova Science
Publishers (2001, New York).

\bibitem{isotr05} V. Baran, M. Colonna, M. Di Toro, et al.,
{\em Phys. Rev.} {\bf C72} (2005) 064620.

\bibitem{tsang92} M.B. Tsang, et al., {\em Phys. Rev. Lett.} {\bf 92},  (2004) 062701.

\bibitem{shiPRC68}
L. Shi, P. Danielewicz, {\em Phys. Rev.} {\bf C68}, (2003) 064604.

\bibitem{BALi} B.A. Li, L.W. Chen,
{\em Phys. Rev.} {\bf C72}, (2005) 064611; L.W. Chen, C.M. Ko, B.A. Li, {\em Phys.
Rev. Lett} {\bf 94}, (2005) 032701.

\bibitem{fabbri04}
 M. Colonna, G. Fabbri, et al.,
{\em Nucl. Phys.} {\bf A742} (2004) 337.

\bibitem{chomazPR}
P.Chomaz, M.Colonna, J.Randrup,
 {\em Phys. Rep.}  {\bf 389} (2004) 263.

 \bibitem{rizzoPRC72}
J. Rizzo, M. Colonna, M. Di Toro, {\em Phys. Rev.} {\bf C72} (2005) 064609.

\bibitem{GalePRC41} C. Gale, G.M. Welke, M. Prakash, et al.,
 {\em Phys .Rev.} {\bf C41} (1990) 1545.

 \bibitem{DanLac02} P. Danielewicz, R. Lacey, W.G. Lynch, {\em Science} {\bf 298} (2002) 1592.


\bibitem{famiano} M.A. Famiano, et al., {\em Phys. Rev. Lett.}
 {\bf 97} (2006) 052701.

\bibitem{danielew} Y. X. Zhang, P. Danielewicz, M. Famiano, et al., arXiv:0708.3684v1 [nucl-th].

\bibitem{mc_DR} M. Colonna, V. Baran, M. Di Toro, H. H. Wolter,  arXiv:0707.3092v1 [nucl-th].

\bibitem{ramimb} F. Rami et al., {\em Phys. Rev. Lett.} {\bf 84} (2000) 1120.

\bibitem{soul04} G.A. Souliotis, M. Velselsky, D.W. Shetty, S.J. Yennello,
 {\em Phys. Lett.} {\bf B588} (2004) 35.

\bibitem{MDT_WCI}  M. Di Toro, A. Olmi, R. Roy,
{\em Eur. Phys. Jour.} {\bf A30} (2006) 65, and refs. therein.

\bibitem{WCI_betty}
M.Colonna and M.B.Tsang, {\em Eur. Phys. J.} {\bf A30} (2006) 165,
and refs. therein.

\bibitem{DeFilipp05}
E. De Filippo, et al., {\em Phys. Rev.}{\bf C71} (2005) 044602 and 064604.

\end{thebibliography}
\end{document}